\documentclass[eqsecnum,showkeys,showpacs,nofootinbib,aps]{revtex4}

\def\beq{\begin{equation}}
\def\eeq{\end{equation}}
\def\bea{\begin{eqnarray}}
\def\eea{\end{eqnarray}}
\def\nn{\nonumber}

\def\pa{\partial}

\def\bpsi{\bar{\psi}}
\def\bdelta{\bar{\delta}}
\def\ol{\overline}
\def\Dt{D_{t}}

\begin{document}
\title{$N=4$ Supersymmetric Quantum Mechanics with Magnetic Monopole}
\author{Soon-Tae Hong}
\email{soonhong@ewha.ac.kr} \affiliation{Department of Science
Education, Ewha Womans University, Seoul 120-750 Korea}
\author{Joohan Lee}
\email{joohan@kerr.uos.ac.kr} \affiliation{Department of Physics,
University of Seoul, Seoul 130-743 Korea}
\author{Tae Hoon Lee}
\email{thlee@ssu.ac.kr} \affiliation{Department of Physics,
Soongsil University, Seoul 156-743 Korea}
\author{Phillial Oh}
\email{ploh@newton.skku.ac.kr} \affiliation{Department of Physics
and Institute of Basic Science, Sungkyunkwan University, Suwon
440-746 Korea}
\date{\today}

\begin{abstract}
We propose an $N=4$ supersymmetric quantum mechanics of a charged
particle on a sphere in the background of Dirac magnetic monopole
and study the system in the $CP(1)$ model approach. By using the
Dirac quantization method, we explicitly calculate the symmetry
algebra taking the operator ordering ambiguity into consideration.
We find that it is given by the superalgebra $su(1\vert 2)\times
su(2)_{\rm rot}$. We also show that the Hamiltonian can be written
in terms of the Casimir invariant of $su(2)_{\rm rot}$ algebra.
Using this relation and analyzing the lower bound for angular
momentum, we find the energy spectrum. We, then, examine the ground
energy sector to find that the $N=4$ supersymmetry is spontaneously
broken to $N=2$ for certain values of the monopole charge.
\end{abstract}
\preprint{hep-th/yymmnn} \maketitle


Since the pioneering work of Dirac the quantum mechanics of a
charged particle in the background of a Dirac magnetic monopole
\cite{dira, cole} has attracted a great deal of attention due to its
rich physical and mathematical properties. Its supersymmetric
extension has been also proposed \cite{hoke} and its various aspects
has been studied \cite{dejo, various, kim, jlee}. In the
supersymmetric version, it was found \cite{dejo} that the system
originally written in $N=1$ formulation \cite{hoke} possesses an
additional (hidden) supersymmetry, making the system in fact $N=2$
supersymmetric. In Ref. \cite{dejo} it was also pointed out that the
hidden supercharge is related to the theory restricted to $S^2$, the
subspace of $R^3$ representing a fixed distance from the location of
the monopole. On the other hand, manifest $N=2$ superspace
formulation of the system is possible on $S^2$ due to its K\"ahler
structure, and the system confined to $S^2$ has been investigated in
detail recently \cite{kim, jlee}. In particular, in Ref. \cite{kim}
the complete energy spectrum with the corresponding wave functions
were found, and in Ref. \cite{jlee} the issue of spontaneous
symmetry breaking (whether or note the ground state is invariant
under the supersymmetries) was discussed using the $CP(1)$ model
approach. In this context, one of the interesting questions would be
to ask whether the supersymmetry of the system could be extended
further.

In this paper, we propose an explicit model of $N=4$ supersymmetric
quantum mechanics of a charged particle on a sphere in the
background of Dirac magnetic monopole. Following the previous work
on $N=2$ supersymmetric quantum mechanics \cite{jlee}, we adopt the
$CP(1)$ model approach. This allows $N=4$ formulation of the system
including the monopole interaction in component formalism. The
dynamical variables consist of bosonic variables $(\bar z_i,~ z_j)$
and fermionic variables $(\bar\psi_{\alpha i}, \psi_{\beta j})$
$(i,j=1,2;~\alpha, \beta=1,2)$ satisfying $\bar{z}\cdot z-1=0,~
\bar{z}\cdot\psi_\alpha=0,~\bar{\psi}_\alpha\cdot z=0.$ Note that
the fermionic variables are doubled in number compared to the $N=2$
case, as indicated by the internal indices, $\alpha, \beta$. The
bosonic variables are related to the space coordinate by the Hopf
map $\vec x = \bar z \vec\sigma z.$

In quantizing the system there appears a parameter associated with
the choice of operator ordering in defining the basic commutation
relations as in the $N=2$ case \cite{jlee}. We study how physical
quantities such as energy and angular momentum depend on this
parameter. After quantization, we find that the symmetry algebra of
our system is given by $su(1\vert 2)\times su(2)_{\rm rot}$. The
bosonic $su(2)$ sector of $su(1\vert 2)$ is the internal rotations
associated with $\alpha, \beta$ indices, whereas the $su(2)_{\rm
rot}$ corresponds the angular momentum. Using the relations
Hamiltonian with the supercharges and the the Casimir invariant of
the rotational algebra we find the energy spectrum. We also
investigate the ground state sector and find that the spontaneous
supersymmetry breaking occurs for some particular values of the
magnetic charge.

 In order to proceed, we  consider $N=4$ supersymmetric
transformations of the following form \beq
\begin{array}{llll}
\vspace{0.2cm} \delta_\alpha z = \psi_\alpha, &\delta_\alpha \bar{z}
= 0, &\delta_\alpha \psi_\beta=0,
&\delta_\alpha \bpsi_\beta=2i\nabla_{\alpha\beta}\bar{z},\\
\bdelta_\alpha z = 0, &\bdelta_\alpha \bar{z} = \bar{\psi}_\alpha,
&\bdelta_\alpha \psi_\beta=2i\nabla_{\alpha\beta} z,
&\bdelta_\alpha \bpsi_\beta=0,
\end{array}\label{rule}\eeq
where  \beq \nabla_{\alpha\beta}\bar{z}=\delta_{\alpha\beta}D_t\bar
z-\frac{i}{2}(\bpsi_\beta
\cdot\psi_\alpha-\delta_{\alpha\beta}\bpsi_\gamma
\cdot\psi_\gamma)\bar{z}, \eeq and its complex conjugate
 \bea
\nabla_{\alpha\beta} z=\delta_{\alpha\beta}D_t
z+\frac{i}{2}(\bpsi_\alpha
\cdot\psi_\beta-\delta_{\alpha\beta}\bpsi_\gamma \cdot\psi_\gamma)
z. \eea  The covariant derivative $D_t$ is defined by
 \bea\Dt
z &=& \left(\pa_{t}-ia\right)z,\nn\\
\Dt\bar{z} &=& \left(\pa_{t}+ia\right)\bar{z},\eea with $a$ given by
\beq a=-\frac{i}{2}(\bar{z}\cdot\dot{z}-\dot{\ol{z}}\cdot
z)-\frac{1}{2}\bpsi_\alpha\cdot\psi_\alpha. \label{aeq} \eeq Note
$a$ is invariant under the supersymmetric transformations
 of Eq. (\ref{rule}).
  One can check that the above supersymmetry transformations
(\ref{rule}) preserve the supersymmetric $CP(1)$ constraints
\cite{adda} \beq \bar{z}\cdot z-1=0,~
\bar{z}\cdot\psi_\alpha=0,~\bar{\psi}_\alpha\cdot z=0.\eeq

Our supersymmetric Lagrangian is then proposed by
 \beq L= 2|\Dt
z|^{2}+\frac{i}{2}(\bpsi_\alpha\cdot\Dt\psi_\alpha-\Dt
\bar{\psi}_\alpha\cdot\psi_\alpha)-\frac{1}{4}((\epsilon_{\alpha\beta}
\bpsi_\alpha \cdot\psi_\beta)^2+ (\bpsi_\alpha
\cdot\psi_\alpha)^2)-2ga, \label{susylag} \eeq which can be
expressed in the following form;
 \bea L&=& 2|\dot{z} -({\bar z}\cdot
\dot{z} )
z|^{2}+\frac{i}{2}(\bpsi_\alpha\cdot\dot{\psi}_\alpha-\dot{\bpsi}_\alpha\cdot\psi_\alpha)
-\frac{i}{2}(\bar{z}\cdot\dot{z}-\dot{\ol{z}}\cdot
z)\bpsi_\alpha\cdot\psi_\alpha\nonumber\\
&-& \frac{1}{4}((\epsilon_{\alpha\beta} \bpsi_\alpha
\cdot\psi_\beta)^2+ (\bpsi_\alpha \cdot\psi_\alpha)^2)+
ig(\bar{z}\cdot\dot{z}-\dot{\ol{z}}\cdot
z-i\bpsi_\alpha\cdot\psi_\alpha). \label{susylagcom}\eea In the
above Lagrangian, we put the electric charge $e=-1$, and $g$ is the
magnetic monopole charge. Compared to the previous $N=2$ case
\cite{jlee}, it has additional quartic fermionic interaction terms
which are essential for the existence of $N=4$ supersymmetry.


Next, we perform the canonical quantization of the system. We
define the momenta $p$ and $\bar{p}$ conjugate,respectively, to
the fields $z$ and $\bar{z}$, \beq
p=2\Dt\bar{z}+\frac{i}{2}(\bpsi_\alpha\cdot\psi_\alpha+2g)\bar{z},~~~
\bar{p}=2\Dt z-\frac{i}{2}(\bpsi_\alpha\cdot\psi_\alpha+2g)z. \eeq
The classical Hamiltonian is \beq H_c=2(\Dt \bar{z})(\Dt
z)-g\bpsi_\alpha\cdot\psi_\alpha+\frac{1}{4}(
 (\epsilon_{\alpha\beta} \bpsi_\alpha
\cdot\psi_\beta)^2 -(\bpsi_\alpha\cdot\psi_\alpha)^{2}).
\label{susyham} \eeq It should be supplemented by the following six
second class constraints \beq C_{1}=\bar{z}\cdot z-1,~~~C_{2}=p\cdot
z+\bar{z}\cdot\bar{p},~~~
C_{3\alpha}=\bar{z}\cdot\psi_\alpha,~~~C_{4\alpha}=\bpsi_\alpha\cdot
z, \label{secconst} \eeq and one first class constraint, \beq
C_{0}=-i(\bar{z}\cdot\bar{p}-p\cdot
z)-\bpsi_\alpha\cdot\psi_\alpha+2g, \label{firconst}\eeq the Gauss
law constraint corresponding to the local $U(1)$ symmetry.
Classically, two quartic terms in the fermion field in Eq.
(\ref{susyham}) cancel each other. We keep it here because they
produce a quadratic term when quantized. We quantize this theory
following the Dirac scheme. We start with the Poisson bracket
relations \beq
\{z_{i},p_{j}\}=\{\bar{z}_{i},\bar{p}_{j}\}=\delta_{ij},~~~
\{\bpsi_{i\alpha},\psi_{j\beta}\}=-i\delta_{ij}\delta_{\alpha\beta},
\eeq with the remaining brackets being zero. To incorporate the
second class constraints, we calculate the Dirac brackets using the
definition given by \beq
\{A,B\}_{D}=\{A,B\}-\{A,C_{a}\}\Theta^{ab}\{C_{b},B\}, \eeq where
$\Theta^{ab}$ is the inverse matrix of
$\Theta_{ab}=\{C_{a},C_{b}\}$. From this result we obtain the
following quantum commutation (and anti-commutation) relations upon
replacing $\{A,B\}_{D}\rightarrow -i[A,B]$, \beq
\begin{array}{ll}
\vspace{0.2cm}
\left[p_{i},z_{j}\right]=-i\delta_{ij}+\frac{i}{2}\bar{z}_{i}z_{j},
&\left[p_{i},\bar{z}_{j}\right]=\frac{i}{2}\bar{z}_{i}\bar{z}_{j},\\
\vspace{0.2cm}
\left[p_{i},p_{j}\right]=\frac{i}{2}(p_{i}\bar{z}_{j}-p_{j}\bar{z}_{i}),
&\left[\bar{p}_{i},p_{j}\right]=\frac{i}{2}(\bar{z}_{j}\bar{p}_{i}-z_{i}p_{j})
-\alpha\psi_{i\alpha}\bpsi_{j\alpha}+\beta\bpsi_{j\alpha}\psi_{i\alpha},\\
\left[\bpsi_{i\alpha},\psi_{j\beta}\right]=\delta_{\alpha\beta}(\delta_{ij}-\bar{z}_{i}z_{j}),
&\left[p_{i},\bpsi_{j\alpha}\right]=i\bpsi_{i\alpha}\bar{z}_{j},
\label{comm}
\end{array}
\eeq with $\alpha+\beta=1$. The above brackets are supplemented by
their Hermitian conjugates, and the remaining commutators all
vanish. They form a straightforward generalization of $N=2$ case
\cite{jlee}. In the second line, the operator ordering in the second
bracket is chosen by the condition that the dynamical variables
commute with the second class constraint, $C_{2}$, ordered as
$p\cdot z+\bar{z}\cdot\bar{p}=0$. Note that this does not fix the
operator ordering completely  and we still have undetermined
$\alpha$ (or $\beta$) in Eq. (\ref{comm}) as in the $N=2$ case
\cite{jlee}.

In order to obtain the symmetry algebra of the system, we compute
the Noether charges associated with various global symmetries. The
space rotations are generated by \beq \left(
\begin{array}{cc}
z_{1}\\
z_{2}
\end{array}
\right)\rightarrow e^{-\frac{i}{2}w^{a}\sigma^{a}} \left(
\begin{array}{cc}
z_{1}\\
z_{2}
\end{array}
\right), \eeq whose operator-ordered conserved charge is given by
\beq
K_{a}=\frac{i}{2}\bar{z}\sigma_{a}\bar{p}-\frac{i}{2}p\sigma_{a}z
+\gamma\bar{z}\sigma_{a}z+\frac{1}{2}\bpsi_\alpha\sigma_{a}\psi_\alpha.
\label{ka} \eeq Here we have added the third term associated with
the operator ordering ambiguity.  We find that $K_{a}$'s generate
the required rotation and satisfy the $SU(2)$ algebra \beq
[K_{a},K_{b}]=i\epsilon_{abc}K_{c},\label{angalg}\eeq provided the
following conditions are satisfied \beq
\alpha=\frac{1}{4}+\gamma,~~\beta=\frac{3}{4}-\gamma.\eeq  The
Noether charge associated with the phase symmetry of the fermionic
variables yields the conserved charge \beq
N_{F}=\bpsi_\alpha\cdot\psi_\alpha. \eeq The supercharges are given
by \beq
Q_\alpha=p\cdot\psi_\alpha,~~~\bar{Q}_\beta=\bpsi_\beta\cdot\bar{p}.\eeq
Note that the supercharges have no ordering ambiguity. The internal
$SU(2)$ rotations define isospin operators $S^{a}$ by \beq
S^{a}=\frac{1}{2}\bpsi_{\alpha}\sigma^{a}_{\alpha\beta}\psi_{\beta}.
\eeq A straightforward calculation yields the following relations
\beq S^2=-\frac{3}{4} N_F^2 + \frac{3}{2} N_F=-\frac{3}{4}
(N_F-1)^2+\frac{3}{4},\label{angrel}\eeq where $S^2= S^aS^a$.
Denoting the eigenvalue of $S^2$ by $s(s+1)$, we find the relations
\bea s=0~~&\leftrightarrow&~~ N_F=0,~~\nonumber\\
s=\frac{1}{2}~~&\leftrightarrow&~~ N_F=1,~~\nonumber\\
s=0~~&\leftrightarrow&~~ N_F=2,~~\label{nf} \eea For local $U(1)$
symmetry we choose as the generator \beq G_0\equiv
-i(\bar{z}\cdot\bar{p}-p\cdot z)-\bpsi_\alpha\cdot\psi_\alpha.
\eeq Classically, this quantity being equal to minus the twice of
the monopole charge is the Gauss law constraints, Eq.
(\ref{firconst}). To use this constraint to select the physical
Hilbert space, however, one should take care of the ordering
problem. This amounts to adding an appropriate constant before
setting $G_0$ equal to $-2g$. As our quantum mechanical
Hamiltonian we choose the following expression \bea H &=&
\frac{1}{2}\left(p\cdot\bar{p}- p\cdot z
\bar{z}\cdot\bar{p}\right)-\frac{i}{2}(\bar{z}\cdot\bar{p}-p\cdot
z)\bpsi_{\alpha}\cdot\psi_{\alpha}-(\alpha+\frac{1}{4})
\bpsi_{\alpha}\cdot\psi_{\alpha}\nonumber\\
&=& \frac{1}{2}\left(p\cdot\bar{p}- p\cdot z
\bar{z}\cdot\bar{p}\right)+\frac{1}{2}\left(G_0-2\alpha+
\frac{1}{2}\right)\bpsi_{\alpha}\cdot\psi_{\alpha}
+\frac{1}{2}(\bpsi_{\alpha}\cdot\psi_{\alpha})^{2}
-\frac{1}{2}\bpsi_{\alpha}\cdot\psi_{\alpha}. \label{quantumham}\eea
One can show that the above Hamiltonian commutes with the
supercharges. It can be obtained by quantizing the classical
Hamiltonian of Eq. (\ref{susyham}) if $G_0-2\alpha+ 1/2$ is
identified with $-2g$. This strongly suggests that the
aforementioned ordering constant is $-2\alpha+\frac{1}{2}$ and
$G_0-2\alpha+\frac{1}{2}$ should be interpreted as the monopole
charge. However, for consistency of this interpretation we must show
that $G_0-2\alpha+\frac{1}{2}$ is quantized according to the Dirac
quantization condition of the magnetic monopole charge. It will be
shown shortly (see Eq. (\ref{kmin}) and discussions below) that this
is indeed the case. Still there is an ambiguity of adding an integer
in the choice of the above mentioned ordering constant. A different
choice of this integer will correspond to a different theory (or a
different interpretation of the theory). In this paper we write the
quantum mechanical Gauss law constraint as \beq
G_0-2\alpha+\frac{1}{2}= -2\tilde{g}, \label{Gauss}\eeq with
$\tilde{g}$ being treated as the monopole charge and our physical
states are required to satisfy the above Gauss law constraint.

A complete algebraic structure of the symmetry generators can be
computed to yield
 \beq
\begin{array}{ll}
\vspace{0.2cm} \left[S^a,S^b\right]=i\epsilon_{abc}S^{c},
&\left[S^a,N_{F}\right]=0,\\
\vspace{0.2cm} \left[Q_{\alpha},N_{F}\right]=Q_{\alpha},
&\left[\bar{Q}_{\alpha},N_{F}\right]=-\bar{Q}_{\alpha},\\
\vspace{0.2cm}
\left[S^a,Q_{\alpha}\right]=-\frac{1}{2}\sigma^{a}_{\alpha\beta}Q_{\beta},
&\left[S^a,\bar{Q}_{\alpha}\right]=\frac{1}{2}\bar{Q}_{\beta}\sigma^{a}_{\beta\alpha},\\
\vspace{0.2cm} \left[Q_{\alpha},Q_{\beta}\right]=0,
&\left[\bar{Q}_{\alpha},\bar{Q}_{\beta}\right]=0,\\
\vspace{0.2cm}
\left[Q_{\alpha},\bar{Q}_{\beta}\right]=2H\delta_{\alpha\beta}-2\tilde{g}
\left(S^a\sigma^{a}_{\alpha\beta}-\frac{1}{2}N_{F}\delta_{\alpha\beta}\right).
\label{susyalg}\end{array} \eeq Also the $K_{a}$ operators in Eq.
(\ref{ka}) satisfy, \beq
\begin{array}{ll}
\vspace{0.2cm} \left[K_{a},K_{b}\right]=i\epsilon_{abc}K_{c},
&\left[K_{a},N_{F}\right]=0,\\
\vspace{0.2cm} \left[K_{a},Q_{\alpha}\right]=0,
&\left[K_{a},\bar{Q}_{\alpha}\right]=0.
\end{array}
\eeq Defining $\tilde N_F\equiv N_F+ 2H/\tilde g$ and
$q_\alpha\equiv Q_\alpha/\sqrt{-2\tilde g}, ~\bar q_\beta\equiv
\bar Q_\alpha/\sqrt{-2\tilde g}~(\tilde g<0)$ we find the last
equation of Eq. (\ref{susyalg}) becomes \beq
\left[q_{\alpha},\bar{q}_{\beta}\right]=-\frac{1}{2}\tilde
N_{F}\delta_{\alpha\beta}+ S^a\sigma^{a}_{\alpha\beta}.\eeq Other
commutation relations remain the same because $[H, S^a]=[H,
N_F]=[H, Q_\alpha]=[H, \tilde Q_\beta]=0$ and  we find that $K_a,
S^a, \tilde N_F, q_\alpha$ and $\bar q_\beta$ generate $su(1\vert
2)\times su(2)_{\rm rot}$ symmetry \cite{marc}. (The case of
$\tilde g>0 $ can be covered by redefining $S^{a}\rightarrow
\frac{1}{2}\bpsi_{\alpha}(-\sigma^{aT}_{\alpha\beta})\psi_{\beta}$,
$N_F\rightarrow -\bpsi_\alpha\cdot\psi_\alpha$ and interchanging
$Q_\alpha\rightarrow \bar Q_\alpha$, $\bar Q_\alpha\rightarrow
Q_\alpha$.)

In order to examine the energy spectrum let us write the Hamiltonian
in terms of the supercharges, as \beq H = \frac{1}{4} [Q_\alpha,
\bar Q_\alpha]_+ -\frac{1}{2}\tilde g N_F,\label{po}\eeq where we
have added the subscript $+$ to emphasize that the bracket is the
anti-commutator. This relation can be obtained by taking the trace
of the last line of Eq. (\ref{susyalg}). Since the first term is
non-negative one finds that the energy is bounded from below by the
second term in the above equation. Note that each sector with a
definite fermion number has a different bound. However, whether or
not these bounds can be saturated depends on the existence of the
states in each sector that are invariant under the supersymmetry. In
other words they may not be the optimal bounds. In order to obtain
the true energy bound for each sector for a given parameter
$\tilde{g}$, we proceed as follows. First, note that \bea
p\cdot\bar{p}- p\cdot z \bar{z}\cdot\bar{p} &=&
p_i(\delta_{ij}-z_i{\bar{z}}_j)\bar{p}_j\nonumber
\\&=& p_i\epsilon_{ij}{\bar z}_j ~ z_l\epsilon_{kl}\bar{p}_k \equiv
a\bar a \eea where we have defined \beq a =
\epsilon_{ij}p_i{\bar{z}}_j, ~~~~ \bar{a} =
\epsilon_{kl}z_l\bar{p}_k.\eeq They satisfy the following
commutation relation \beq [a,\bar{a}] =
-G_0+2\alpha+\frac{3}{2}-2\bpsi_{\alpha}\cdot\psi_{\alpha}=
2(\tilde{g}-\Sigma),\eeq where $\Sigma\equiv N_F-1$ can be
regarded as the total spin along the radial direction, and Eq.
(\ref{Gauss}) was used. Eigenvalues of the spin operator,
$\Sigma$, consists of three values, $\sigma=+1,0,-1$, reflecting
that we have two spin half degrees of freedom. Next, we need the
bound for $a\bar{a}$. If $\tilde{g}-\sigma\le 0$, the bound is
zero, and the state saturating this bound is obtained by imposing
$\bar{a}=0$. If $\tilde{g}-\sigma\ge 0$, $a\bar{a}$ is bounded by
$2(\tilde{g}-\sigma)$ as one can see from
$a\bar{a}=\bar{a}a+2(\tilde{g}-\sigma)$ and $\bar{a}a\ge 0$. In
the latter case the bound is saturated by the states satisfying
$a=0$. These two cases can be combined into \beq a\bar{a}\ge
|\tilde{g}-\sigma|+(\tilde{g}-\sigma). \eeq This result together
with Eqs. (\ref{quantumham}), (\ref{Gauss}) yields the bound for
the energy \bea E_{\rm min} &=&
\frac{1}{2}\left(|\tilde{g}-\sigma|+(\tilde{g}-\sigma)\right)-\tilde{g}(\sigma+1)
+\frac{1}{2}(\sigma+1)^2- \frac{1}{2}(\sigma+1)\\
&=& \frac{1}{2}|\tilde{g}-\sigma|\left(|\tilde{g}-\sigma|+1\right)-
\frac{1}{2}\tilde g \left(\tilde g+1\right).\label{emin}\eea By
comparing this minimum energy with $-\frac{1}{2}\tilde g N_F$ of Eq.
(\ref{po}) we conclude that the following types of supersymmetric
ground states exist: $\sigma=-1$ if $\tilde g\le-1$, $\sigma=0$ if
$\tilde g=0$, and $\sigma=1$ if $\tilde g\ge 1$.

One can obtain further information from the relation between the
Hamiltonian and the angular momentum squared \beq
H=\frac{1}{2}\left(K^2-\tilde g \left(\tilde g+1\right)\right).\eeq
Comparing this with Eq. (\ref{emin}) leads to the minimum value of
the angular momentum quantum number \beq k_{\rm
min}=|\tilde{g}-\sigma|. \label{kmin} \eeq Since the angular
momentum quantum number must be half-integer or integer and the spin
$\sigma$ is an integer, $\tilde g$ must also be a half-integer or an
integer. In particular, one finds that $k_{\rm min}$ is a
half-integer if $\tilde g$ is a half-integer, and an integer if
$\tilde g$ is an integer. This confirms that the quantization of
$\tilde g$  and $\tilde g$ can indeed be interpreted as the
effective monopole charge. More generally, the angular momentum
quantum number can be written as \beq k=k_{\rm min}+n,~~
~~(n=0,1,2,\cdots) \eeq and the energy spectrum can be written as
\beq E=\frac{1}{2}k(k+1)-\frac{1}{2}\tilde g \left(\tilde
g+1\right).\eeq We give the diagram for $k$ versus $\tilde{g}$ in
Fig. 1. Among the states satisfying Eq. (\ref{kmin}), the ground
states are represented by the points marked with dots and circles.
The dots correspond to the supersymmetric ground states for given
values of $\tilde g$. The cases, $\tilde{g} = \pm 1/2$, are somewhat
special. In these cases there is no $N=4$ supersymmetric states. In
other words, there is no state killed by both $Q_\alpha$ and
$\bar{Q}_\alpha$. For $\tilde{g}=1/2$, for instance, the ground
states, marked with a circle in Fig. 1, consist of two sectors,
$(k=1/2, \sigma=0$) and $(k=1/2, \sigma=1)$, each of which has a
twofold angular momentum degeneracy. However, one can argue in this
case that the ground state still has $N=2$ supersymmetry left over.
Note that the supercharges commute with the Hamiltonian and the
angular momentum. Thus, applying the supercharges to any state will
not change the energy and angular momentum quantum numbers. On the
other hand the commutation relations $[\Sigma, Q_\alpha]=-Q_\alpha$
and $[\Sigma,\bar{Q}_\alpha]=\bar{Q}_\alpha$ tell us that $Q_\alpha,
\bar Q_\alpha$ play the role of lowering and raising operators of
the fermion number. Now, let $|\psi\rangle$ be a ground state with
$\sigma=0$. Then $Q_\alpha|\psi\rangle=0$ because there is no
$\sigma=-1$ state to be mapped in the ground energy sector.
Similarly, $\bar{Q}_\alpha|\psi\rangle$ belongs to the $\sigma=1$
sector of the lowest energy level. Simple counting together with the
fact that the supercharges commute with $K_a$ suggests that a
certain linear combination of $\bar{Q}_\alpha|\psi\rangle$ should
vanish. To conclude, for $\tilde{g} = \pm 1/2$ the ground state is
invariant under the half of the supersymmetry.

\begin{figure}[t]
\setlength{\unitlength}{1.5cm}
\begin{center}
\begin{picture}(6.5,6.5)(0,0)
\put(7,1.95){$\tilde{g}$} \put(3.1,5.3){$k$}
\put(-0.52,4.55){$\sigma=-1$} \put(0.42,4.55){$\sigma=0$}
\put(1.22,4.55){$\sigma=1$}\put(4.5,4.55){$\sigma=-1$}
\put(5.5,4.55){$\sigma=0$} \put(6.3,4.55){$\sigma=1$}
\put(3.24,1.16){$-1$} \put(3.22,2.85){1} \put(3.22,3.65){2}
\put(3.22,4.45){3} \put(3.95,1.8){1} \put(4.75,1.8){2}
\put(5.55,1.8){3}\put(6.35,1.8){4} \put(3.25,1.8){0}
\put(2.2,1.8){$-1$} \put(1.4,1.8){$-2$} \put(0.6,1.8){$-3$}
\put(-0.2,1.8){$-4$} \put(3.15,0.8){\line(1,0){0.1}}
\put(3.15,1.2){\line(1,0){0.1}} \put(3.15,1.6){\line(1,0){0.1}}
\put(3.6,1.95){\line(0,1){0.1}} \put(4.2,1.95){\line(0,1){0.1}}
\put(4.4,1.95){\line(0,1){0.1}} \put(4.8,1.95){\line(0,1){0.1}}
\put(5.2,1.95){\line(0,1){0.1}} \put(5.6,1.95){\line(0,1){0.1}}
\put(6.0,1.95){\line(0,1){0.1}} \put(6.4,1.95){\line(0,1){0.1}}
\put(0.4,1.95){\line(0,1){0.1}} \put(1.2,1.95){\line(0,1){0.1}}
\put(2.0,1.95){\line(0,1){0.1}} \put(2.4,1.95){\line(0,1){0.1}}
\put(2.8,1.95){\line(0,1){0.1}} \put(1.6,1.95){\line(0,1){0.1}}
\put(0.8,1.95){\line(0,1){0.1}} \put(0.0,1.95){\line(0,1){0.1}}
\put(4.0,1.95){\line(0,1){0.1}} \put(-0.1,2){\vector(1,0){6.8}}
\put(3.2,0.5){\vector(0,1){4.5}} \thicklines
\put(2.4,2){\line(1,0){1.6}} \put(4.0,2){\circle*{0.1}}
\put(3.2,2){\circle*{0.1}} \put(2.4,2){\circle*{0.1}}
\put(2.0,2.4){\line(1,0){2.4}} \put(2.0,2.4){\circle*{0.1}}
\put(2.8,2.4){\circle{0.1}} \put(3.6,2.4){\circle{0.1}}
\put(4.4,2.4){\circle*{0.1}} \put(1.6,2.8){\line(1,0){3.2}}
\put(1.6,2.8){\circle*{0.1}} \put(4.8,2.8){\circle*{0.1}}
\put(1.2,3.2){\line(1,0){4}} \put(1.2,3.2){\circle*{0.1}}
\put(5.2,3.2){\circle*{0.1}} \put(0.8,3.6){\line(1,0){4.8}}
\put(0.8,3.6){\circle*{0.1}} \put(5.6,3.6){\circle*{0.1}}
\put(0.4,4){\line(1,0){5.6}} \put(0.4,4){\circle*{0.1}}
\put(6.0,4){\circle*{0.1}} \put(0.0,4.4){\line(1,0){6.4}}
\put(0.0,4.4){\circle*{0.1}} \put(6.4,4.4){\circle*{0.1}}
\thinlines \put(3.2,2){\line(1,1){2.5}}
\put(2.4,2){\line(-1,1){2.5}} \put(4.0,2){\line(1,1){2.5}}
\put(3.2,2){\line(-1,1){2.5}} \put(2.4,2){\line(1,1){2.5}}
\put(4.0,2){\line(-1,1){2.5}}
\end{picture}
\end{center}
\caption{Diagram for $k$ versus $\tilde{g}$.}
\end{figure}
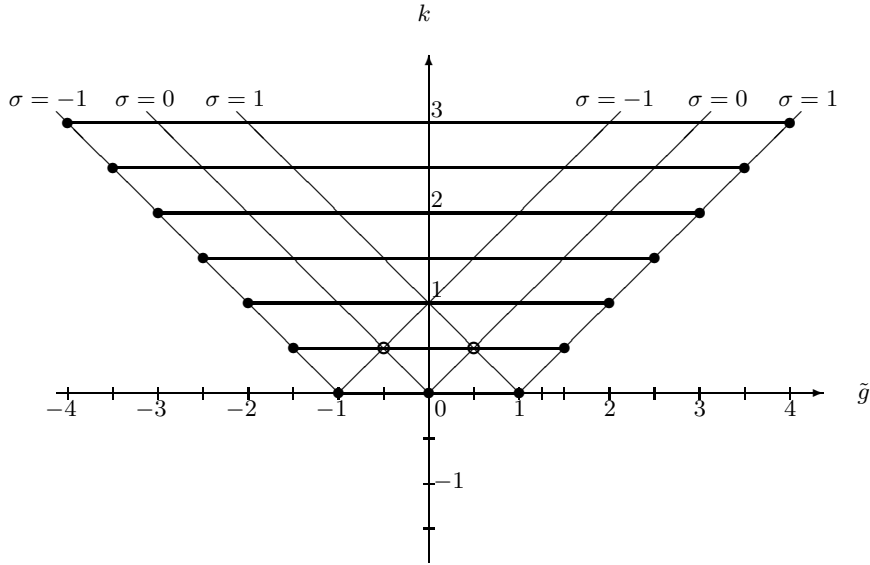

In summary, we have shown that  the quantum mechanics of a charged
particle on a sphere in the background of Dirac magnetic monopole
allows $N=4$ supersymmetric extension. Using the Dirac quantization
procedure, we explicitly calculated the symmetry algebra and found
that it is given by the superalgebra $su(1\vert 2)\times su(2)_{\rm
rot}$. We also investigate the spectrum of the Hamiltonian which is
given by the Casimir invariant of the
 $su(2)_{\rm rot}$ symmetry. By analyzing the ground energy sector
 we found that the supersymmetry is spontaneously broken for
 particular values of $\tilde g= \pm
 \frac{1}{2}.$
There are a few unresolved aspects deserving a further study: The
first one is to look for a manifest superspace formulation of our
system by using the $N=4$ chiral superfield. The other is the
construction of the complete wavefunctions by explicitly realizing
Eq. (\ref{comm}) as differential operators on an appropriate
function space. Some of these are currently under progress.


\acknowledgments
STH would like to acknowledge financial support in
part from the Korea Science and Engineering Foundation grant
(R01-2000-00015). PO was supported by Korea Research Foundation grant
(R05-2004-000-10682-0).  The authors would like to thank the ATCTP
for the hospitality during their visit.

\end{document}